\documentclass[12pt,preprint]{aastex}         
\usepackage{graphicx}
\usepackage{epstopdf}
\def\be{\begin{equation}}
\def\ee{\end{equation}}

\def\lambar{\lambda\llap {--}}

\def\lambar{\lambda\llap {--}}

\def\lsim{\lower 2pt \hbox{$\, \buildrel {\scriptstyle <}\over
         {\scriptstyle \sim}\,$}}

\begin{document}
\newcommand{\figureout}[2]{ \figcaption[#1]{#2} }       

\title{The Vela Pulsar: Results from the First Year of {\it Fermi} LAT Observations}
\author{
A.~A.~Abdo\altaffilmark{2,3}, 
M.~Ackermann\altaffilmark{4}, 
M.~Ajello\altaffilmark{4}, 
A.~Allafort\altaffilmark{4}, 
W.~B.~Atwood\altaffilmark{5}, 
L.~Baldini\altaffilmark{6}, 
J.~Ballet\altaffilmark{7}, 
G.~Barbiellini\altaffilmark{8,9}, 
M.~G.~Baring\altaffilmark{10}, 
D.~Bastieri\altaffilmark{11,12}
% ...
}
% ...
\altaffiltext{2}{Space Science Division, Naval Research Laboratory, Washington, DC 20375, USA}
\altaffiltext{3}{National Research Council Research Associate, National Academy of Sciences, Washington, DC 20001, USA}
\altaffiltext{4}{W. W. Hansen Experimental Physics Laboratory, Kavli Institute for Particle Astrophysics and Cosmology, Department of Physics and SLAC National Accelerator Laboratory, Stanford University, Stanford, CA 94305, USA}
\altaffiltext{5}{Santa Cruz Institute for Particle Physics, Department of Physics and Department of Astronomy and Astrophysics, University of California at Santa Cruz, Santa Cruz, CA 95064, USA}
\altaffiltext{6}{Istituto Nazionale di Fisica Nucleare, Sezione di Pisa, I-56127 Pisa, Italy}
\altaffiltext{7}{Laboratoire AIM, CEA-IRFU/CNRS/Universit\'e Paris Diderot, Service d'Astrophysique, CEA Saclay, 91191 Gif sur Yvette, France}
\altaffiltext{8}{Istituto Nazionale di Fisica Nucleare, Sezione di Trieste, I-34127 Trieste, Italy}
\altaffiltext{9}{Dipartimento di Fisica, Universit\`a di Trieste, I-34127 Trieste, Italy}
\altaffiltext{10}{Rice University, Department of Physics and Astronomy, MS-108, P. O. Box 1892, Houston, TX 77251, USA}
\altaffiltext{11}{Istituto Nazionale di Fisica Nucleare, Sezione di Padova, I-35131 Padova, Italy}
\altaffiltext{12}{Dipartimento di Fisica ``G. Galilei", Universit\`a di Padova, I-35131 Padova, Italy}
% ...

\begin{abstract}
We report on analysis of timing and spectroscopy of the Vela pulsar using eleven months of observations 
with the Large Area Telescope on the {\it Fermi} Gamma-Ray Space Telescope. The intrinsic brightness of Vela at GeV energies 
combined with the angular resolution and sensitivity of the LAT allow us to make the most detailed study 
to date of the energy-dependent light curves and phase-resolved spectra, using a LAT-derived timing model.  
The light curve consists of two peaks (P1 and P2) connected by bridge emission containing a third peak (P3).
We have confirmed the strong decrease
of the P1/P2 ratio with increasing energy seen with EGRET and previous {\it Fermi} LAT data, and observe that 
P1 disappears above 20 GeV.  The increase with energy
of the mean phase of the P3 component can be followed with much
greater detail, showing that P3 and P2 are present up to the highest energies of pulsation.
We find significant pulsed emission at phases outside the main profile, indicating that magnetospheric
emission exists over 80\% of the pulsar period.
With increased high-energy counts the phase-averaged spectrum is seen to depart from a power-law with simple 
exponential cutoff, and is better fit with a more gradual cutoff.  The spectra in fixed-count phase bins 
are well fit with power-laws with exponential cutoffs, revealing a strong and complex phase dependence of 
the cutoff energy, especially in the peaks.  By combining these results with predictions of the outer 
magnetosphere models that map emission characteristics to phase, it will be possible to probe the particle acceleration
and the structure of the pulsar magnetosphere with unprecedented detail. 
\end{abstract} 


\pagebreak
  
\section{Introduction}

The Vela pulsar is the brightest non-flaring source in the GeV $\gamma$-ray sky and therefore offers the best hope to reveal
the inner workings of the pulsar accelerator.  Pulsars are the most luminous high-energy Galactic $\gamma$-ray sources, 
primarily because they radiate the bulk of their spin-down power in the GeV band.  But as they reach
their highest efficiency levels, their spectra cut off exponentially at GeV energies.  Thus, 
pair-conversion telescopes such as the {\it Fermi} LAT, with their prime 
sensitivity around 0.1 - 10 GeV, are the best instruments with which to study the pulsar machine.  
Being the first and best target, 
Vela has a long history of $\gamma$-ray observations beginning with the first 
detection of high-energy pulsations by SAS-2 (Thompson et al. 1975) followed by the first phase-resolved studies
with COS-B (Grenier, Hermsen \& Clear 1988) and 
EGRET (Kanbach et al. 1994, Fierro et al. 1998).  Naturally, Vela was the first source to be studied by 
AGILE (Pellizzoni et al. 2008), 
and the {\it Fermi} LAT, which used Vela for preformance tuning during the month-long commissioning phase following
its launch on June 11, 2008.  A middle-aged pulsar, with period $P = 0.089$ s, period derivative $\dot P = 1.24 
\times 10^{-13}\,\rm s\,s^{-1}$, characteristic age $\tau = 12$ kyr and spin-down power $\dot E_{\rm sd} = 6.3 
\times 10^{36}\,\rm erg\,s^{-1}$, Vela is not the most energetic of the known $\gamma$-ray pulsars, but it is
one of the closest to Earth at $d = 287^{+19}_{-17}$ pc (Dodson et al. 2003).

Most of the models for pulsed emission from rotation-powered pulsars like Vela assume an origin inside the magnetosphere from  
charged particles accelerating along open magnetic field lines (those that do not close within the light cylinder).
There are a few models that place the site of pulsed emission outside the light cylinder in the wind zone (Petri \& Kirk 2005).
The magnetospheric models have divided into two main types: polar cap models (e.g. Daugherty \& Harding 1996) 
where the $\gamma$-ray emission comes from pair cascades
near the neutron star surface, and outer magnetosphere models where the $\gamma$-ray emission comes from outer gaps (e.g. 
Romani \& Yadogaroglu 1995, Cheng, Ruderman \& Zhang 2000, Hirotani 2008) or from slot gaps (Muslimov \& Harding 2004, 
Harding et al. 2008).  Although all of these models can produce Vela-like light curves, they can be distinguished by
the required inclination and viewing angles, by the shape of the spectral cutoffs, by the phase of the $\gamma$-ray peaks
relative to the radio pulse and by the phase-resolved spectra.

Previous studies of Vela $\gamma$-ray pulsations by SAS-2, EGRET, AGILE and {\it Fermi} have revealed a light curve with 
two narrow and widely separated peaks (P1 and P2, separated by 0.4 in phase), neither of which is in phase with the single radio pulse.  There
is also complex bridge emission containing a third broader peak (P3) between the two main $\gamma$-ray peaks.  
{\it Fermi} LAT observations (Abdo et al. 2009a) showed that P3 is a distinct component in the light curve which 
moves to larger phase with increasing energy while the two main peaks remain at constant phase.  These early {\it Fermi}
observations also found that the main peaks are very sharp, with the second peak having a slow rise and fast decay.
The phase-averaged spectrum was fit above 200 MeV with a power law plus super-exponential cutoff and a cutoff shape 
sharper than a simple exponential was rejected with a significance of 16$\sigma$.  This result thus rules out near-surface
emission, as proposed in polar cap cascade models (Daugherty \& Harding 1996), which would exhibit a sharp spectral cutoff
due to magnetic pair-production attenuation.  

The {\it Fermi} LAT has now collected data since August 4, 2008 in sky-survey mode, observing the entire sky every three hours.
These observations have increased the photon statistics for Vela by a factor of more than 5 over those of Abdo et al. (2009a),
allowing a much deeper level of analysis.  In addition, we are able to use a purely LAT-derived timing solution for the first 
time, which gives smaller RMS residuals than the radio ephemeris. 
This paper reports the results and implications of this analysis, starting in Section 2 with a description of the 
$\gamma$- ray observations, followed by the LAT timing solution for Vela.  Section 3 presents the results on the
light curve at different energies, as well as the phase-averaged and phase-resolved spectra and 
results on variability of flux, while Section 4 presents some implications and possible use of the 
results for probing the geometry and energetics of high-energy emission and particle acceleration.

\section{Observations}

\subsection{{\it Fermi} LAT observations} \label{LATobs}

The LAT, the main instrument on {\it Fermi}, is a pair-production
telescope sensitive to $\gamma$ rays from 20 MeV to $> 300$ GeV with on-axis 
effective area $> 1$ GeV of $\sim 8000\, \rm cm^2$, exceeding that of EGRET by a factor of about five.  
The LAT is made of a high-resolution silicon tracker, a hodoscopic CsI electromagnetic calorimeter
and an anticoincidence detector for charged particles background identification. The full
description of the instrument and its performance can be found in Atwood et al. (2009).  The LAT field-of-view ($\sim
2.4$ sr) covers the entire sky every three hours (2 orbits). The single-event point spread function
(PSF) strongly depends on both the energy and the conversion point in the tracker, but
less on the incidence angle. For 1 GeV normal incidence conversions in the upper section
of the tracker the PSF 68\% containment radius is 0\fdg 6. Timing is provided to the LAT 
by the satellite GPS clock and photons are timestamped to an accuracy better than 300 ns.  
The improved detection capabilities
of the LAT combined with its observing strategy lead to an increment of a factor $\sim 30$ in
sensitivity with respect to its predecessor EGRET.

For results presented here, we used data collected starting August 4, 2008 and extending 
until July 4, 2009.  We select photons in the event class with strongest background rejection (`Diffuse' class) 
that are within a radius of 15 degrees of 
the pulsar position and excluded periods when the pulsar was viewed at zenith angles $> 105^{\circ}$ to 
minimize contamination by photons generated by cosmic ray interactions in the Earth's atmosphere.  
We use photons between 0.1 and 300 GeV for the spectral analysis, due to the large systematic errors 
in the LAT effective area for energies below 0.1 GeV (Abdo et al. 2009e).   We choose photons with energy 
between 0.02 and 300 GeV for the timing analysis, since it is 
not strongly affected by the effective area issues, in order 
to increase the statistics and examine the low energy light curves.
For the timing analysis, the total number of events above 20 MeV, in the region-of-interest 
(ROI) selection defined in Abdo et al. (2009a), is 152119, 
and for the spectral analysis, the number of photons above 100 MeV and inside a ROI (see
\S \ref{sec:lc}) of $15^\circ$ radius is 471324.

\subsection{{\it Fermi} LAT timing solution} \label{LATtime}

In order to do phase-resolved spectroscopy and analysis of the very sharp features of the pulse profile using 11 months of LAT data, we require a timing model for the pulsar that allows us to compute the pulse phase for each detected photon. This timing model must be valid for the full observation interval and should have an accuracy better than the finest time binning used in the analysis.  For the timing model we bin the light curve into variable width bins with $\sim1000$ photons/bin.  This results in a minimum bin width of 128 $\mu$s.  We thus require the timing model to have an accuracy better than this, so that errors in the model do not dominate the apparent width of the light curve features.

The Vela pulsar is routinely timed at Parkes Radio Telescope (Weltevrede et al. 2009) and we obtained a long-term timing model from those observations.  However for this work, we choose to use a timing model derived purely from LAT observations.  The LAT data, which are taken in survey mode, are well suited for constructing regular time of arrival (TOA) measurements and the LAT achieves a very high signal to noise ratio on the Vela pulsar.  In a 2-week integration, we are able to achieve a TOA measurement error of $\sim 40 \mu$s. 

We measured 6 TOAs, spaced at 5-day intervals over the commissioning phase of the mission (2008 June 25 through August 4), and 24 TOAs, spaced at 2-week intervals over the survey portion of the mission (2008 August 4 through 2009 July 15). The TOAs were computed using an unbinned maximum likelihood method as described in Ray et al. (2009).  The template used was an empirical Fourier decomposition measured from the full mission light curve.  The TOAs were fitted to a timing model using \textsc{Tempo2} (Hobbs et al. 2006).  The model uses a position, proper motion, and parallax determined from the radio observations (Dodson et al. 2003) and all were held fixed.  The free parameters in the model are the pulsar frequency and first two frequency derivatives, as well as 3 sinusoidal \texttt{WAVE} terms (Edwards et al. 2006) to model the strong timing noise apparent in this pulsar.  We define phase zero for the model based on the fiducial point from the radio timing observations, which is the peak of the radio pulse at 1.4 GHz. The RMS residuals of the TOAs with respect to the fitted model is 63 $\mu$s, which is fully adequate for the present analysis. The complete timing model used for this analysis will be made available on the {\it Fermi} Science Support Center (FSSC) website\footnote{http://fermi.gsfc.nasa.gov/ssc/data/access/lat/ephems/}.  Phases are computed for the LAT data using the \texttt{fermi} plugin provided by the LAT team and distributed with \textsc{Tempo2}.

\section{Results}

\subsection{Light curves} \label{sec:lc}

To generate the light curves we used an additional selection for photons of energy $E_{\rm GeV}$ in GeV 
within an angle $\theta <$ max$[1.6 - 3\log_{10}(E_{\rm GeV}),1.3]$ degrees of the pulsar position (Abdo et al. 2009a).    
This gives a larger number of photons at high energy relative to the number of photons within the LAT point spread 
function at that energy.   Using this selection and the set of cuts described in 
Section \ref{LATobs}, we have a total number of 152119 photons.  We determine the number of background events to 
be 25100 using the background percentage 16.5\% estimated from the simulation described in \S 3.2.2, 
which leaves 127000 photons above background.
Since we used the {\it Fermi} LAT timing solution described in Section \ref{LATtime}, we corrected the photon 
arrival times to the geocenter 
(and applied an additional correction for the position of {\it Fermi} relative to the geocenter) before phase-folding the photons.  
Figure 1 shows the light curve for the full energy band 0.02 - 300 GeV, displayed in fixed count phase bins to 
exhibit the fine features in detail.  The Poisson error on the light curve is 3.65\% of each bin content. 
The minimum width phase bin is 0.000753 in phase which corresponds to $67 \mu$sec (the timing rms is 63 $\mu$sec). 
The two main peaks are both quite narrow, but also show distinct differences (as seen in 
the insets).  The first peak (P1) is nearly symmetric in its rise and fall, while the second peak (P2) is very asymmetric, 
with a slow rise and more rapid fall.  We fit P1, P2 and P3 jointly with asymmetric Lorentzian functions
\begin{eqnarray}  \label{eqn:Peakfit}
\cal{L}({\it x}) & = A_0 / \{{1+ [(x - x_0)/{\rm HWHM1}]^2}\}, &  x < x_0 \\
           & = A_0 / \{{1+ [(x - x_0)/{\rm HWHM2}]^2}\}, & x > x_0 \nonumber
\end{eqnarray}
for P1 and P2, where $A_0$, $x_0$, ${\rm HWHM1}$ and ${\rm HWHM2}$ are the free parameters for amplitude, location, and half-width-at-half maximum for the leading edge (left) and trailing edge (right) of the peaks, and a log-normal function for P3.  The log-normal function was chosen to fit P3 over a Lorentzian or Gaussian function.  A comparison of similar joint fits of P1, P2 and P3 asssuming the other functions for P3 resulted in significantly larger $\chi^2$ for Gaussian and Lorentzian functions in all energy bands except for 3-8 GeV and 8-20 GeV bands, where the $\chi^2$ are comparable for log-normal and Gaussian.  The parameters of the light-curve fit to the function in Eqn (\ref{eqn:Peakfit}) are given in the last column of Table 1.  It is apparent that the outer half-widths of the two peaks are comparable, but the inner width of P2 is significantly larger than the P1 inner width.  The off-pulse region between phase 0.7 and 1.0 was analyzed by Abdo et al. (2009b) and shows evidence for extended emission from the Vela X pulsar wind 
nebula.  

Figure 2 displays the light curves in seven exclusive energy bands, showing the dramatic changes in the different components
with energy.  We find 35 events in the highest energy band, above 20 GeV, and the highest energy event within the 95\%
containment radius is at 44 GeV. 
The narrowing of P1 and P2, as well as the movement of P3 to higher phase with increasing energy are apparent.
To quantify the energy dependence of the light curve, we have jointly fit two-sided Lorentzian peaks to P1 and P2, 
and a log-normal fit to P3 as a function of energy.  The results are shown in Table 1 and Figures 3 and 4.  
The positions of P1 and P2 are constant with energy within our measurement errors, and the phase of the P3 
centroid shows a marked increase with increasing energy as first noted by Abdo et al. (2009a).  
Both P1 and P2 widths, shown in Figure 3, decrease with increasing energy, but we find that this decrease 
is caused primarily by a decrease in the width 
of the outer wings of the peaks.  The peak inner widths show a more complex energy evolution, initially increasing with energy,
but decreasing sharply at the highest energies.  In the case of P2, the broadening followed by a sudden narrowing of the 
inner wing may be caused
by the splitting off of an additional feature at energy $> 8$ GeV.   This feature can indeed be seen in the $>$
8 GeV light curve in Figure 2.  The width of P3 is found to also decrease
with increasing energy above 100 MeV, as shown  in Figure 4.  The ratio of P1/P2 heights decreases with energy, in agreement with Abdo et al. (2009a), with P1 disappearing above about 20 GeV. 
Thus P3 and P2 are the only remaining features in the highest energy light curve. 
Figure 5 displays the energy evolution of the light curve in a 2-D plot of pulse profile vs. energy, with 20 energy bands along the y-axis equally spaced in Log(E) and 100 equal-width phase bins along the x-axis covering one pulse period. In order to highlight the structures in each energy band with the same scale, we have normalized the total number of events in one full pulse period in each energy band to 100. The color scale shows the relative counts in each bin on a linear scale.  
This type of view clearly shows the movement of P3 with phase and establishes 
it as an independent component of the Vela light curve.  The possible appearance of a spur on the P2 inner edge above 8 GeV can also be seen.

In Figure 6, we display a zoomed-in view of the off-pulse 
phase region in three different energy bands.  The extension of the trailing edge of P2 out to a phase around 0.8 is
clear, especially in the low energy band.  However, in the $> 1$ GeV band the trailing edge of P2 has dropped to background at phase around 0.65.  The energy dependence of the P1 and P2 widths, noted in Figure 3, thus extends to the peak trailing edges,
but more strongly for P1.  
The dashed vertical lines mark the phases of the RXTE P3 ($0.87 \pm 0.02$) and P4 ($1.006 \pm 0.004$)
(Harding et al. 2002).  P4 of the RXTE light curve is one of the few high energy peaks at the radio peak phase, but we see
no significant enhancement of flux at this phase in the {\it Fermi} light curve.  
AGILE has reported a marginal detection of a fourth peak at the $4 \sigma$ level in the Vela light curve at a phase of 
approximately 0.9 in the 0.03 - 0.1 GeV 
band (Pellizzoni et al. 2009), consistent with the location of RXTE-P3.  
We do not detect any significant feature above background in the LAT light curve at or near phase 
0.9 in our lowest energy band (0.02 - 0.1 GeV), or indeed in any energy band.

\subsection{Spectra}

With the large number of counts from Vela in the eleven months of {\it Fermi} observations, we can study the phase-averaged spectrum of
the pulsar in much more detail than was possible after only the first few months.  In particular, the statistics at the highest energies
are greatly improved, so that the shape of the spectral cutoff can be scrutinized more accurately.  In addition, it is now possible to
measure the parameters of the phase-resolved spectrum by performing fits of a power law with exponential cutoff in small phase intervals.

\subsubsection{Phase-averaged spectrum}

A binned maximum likelihood fit was performed to study the phase-averaged  pulsar spectrum using the spectral fitting tool \emph{gtlike}, version v9r15p2 of the LAT Science Tools and the P6\_V3 Instrument Response Function.  
The analysis was done on a $20^{\circ}\times20^{\circ}$ region (a square region that inscribed a circle of radius $10^{\circ}$) centered on the radio position of the pulsar with 10 bins per decade in energy.  All point sources within $15^{\circ}$ of the pulsar, found above the background with a significance $\geq5\sigma$, were included in the model in addition to a uniform disk with extension $0.88^\circ$ for the Vela X remnant.  Spectral parameters for point sources $>10^{\circ}$ from the pulsar were kept fixed to the values found in a preliminary version of the year one LAT catalog 
(Abdo et al. 2009d). The Galactic diffuse background was modeled using the \emph{gll\_iem\_v02} map cube available from the FSSC.  The extragalactic diffuse and residual instrument backgrounds were modeled jointly using the \emph{isotropic\_iem\_v02} template, also available from the FSSC.   An off-pulse region, defined as $\phi \in[0.8,1.0]$, was selected and a fit was done without the pulsar.  The source model was then adjusted to the exposure-corrected, off-pulse values for a phase-averaged fit in which the spectral parameters for all sources $\leq10^{\circ}$ of the pulsar were kept free.  The pulsar was fit assuming a power law plus hyper-exponential cutoff spectral model of the form
\be  \label{hexp}
{dN(E)\over dE} = A E^{-\Gamma}\,\exp[-(E/E_c)^b]
\ee

\noindent and allowing the parameter $b$ to vary in the fit.  We find that the spectrum is best fit with a value of $b < 1$, as 
found by Abdo et al.(2009a).  The Vela phase-averaged spectrum thus appears to be cutting off more slowly than a simple exponential.  The spectral energy distribution with the best fit parameters is shown in Figure 7, along with points derived from \emph{gtlike} fits to individual energy intervals in which the pulsar was fit with a power 
law spectral model.  The best fit parameters for the phase-averaged
spectrum are $A = (3.63 \pm 0.25 \pm 1.01) \times 10^{-9}\,\rm cm^{-2}\,s^{-1}\,MeV^{-1}$, $\Gamma = 1.38 \pm 0.02 + 
^{+0.07}_{-0.03}$, 
$E_c = 1.36 \pm 0.15 +^{+1.0}_{-0.5}$ GeV and $b = 0.69 \pm 0.02 + ^{+0.18}_{-0.10}$, where the first errors are statistical 
and the second are systematic.  As will be discussed in \S \ref{sec:phase_res}, the preference for $b < 1$ in the phase-averaged spectral fit is expected from the large variation in $E_c$ with phase of the phase-resolved spectra.  
To derive the systematic errors on the fits parameters, we used bracketing instrument response functions that propagate the
effective area uncertainty to uncertainties in the parameters.\footnote{The systematic uncertainties were estimated by applying the same fitting procedures described above and comparing results using bracketing instrument response functions (IRFs) which assume a systematic uncertainty in the effective area of $\pm$10\% at 0.1 GeV, $\pm$5\%  near 0.5 GeV, and $\pm$20\% at 10 GeV with linear extrapolations, in log space, between.  To further address the specific systematics associated with an exponentially cutoff spectrum, these systematics were multiplied by a fractional uncertainty of $\pm$1 at 0.1 GeV, $\mp$1 near 3 GeV and $\pm$1 for energies above 10 GeV with linear extrapolations, in log space, between.  The resulting correction factor is used to perturb the effective area from that defined in the P6\_V3 IRFs.}
Since there is significant degeneracy between the $\Gamma$, $E_c$ and $b$ 
parameters in the fit, the derived systematic errors are asymmetric.
Comparing the log(likelihood) values from this fit with 
the same fit holding $b$ fixed to 1, we can exclude a simple exponential fit at about the $11\sigma$ level.
We also tried fitting the spectrum with a power law with simple exponential cutoff plus
an additional power law at high energy, but this form gave a fit with lower significance.  
The phase-averaged integral photon flux in the range
0.1 - 100 GeV is found to be $F(0.1 - 100 \rm GeV) = (1.070 \pm 0.008 \pm 0.030) \times 10^{-5}\,\rm ph\,cm^{-2}\,s^{-1}$.  
The energy flux in this spectral range is $H(0.1 - 100 \rm GeV) = (8.86 \pm 0.05 \pm 0.18) \times 10^{-9}
\,\rm erg\,cm^{-2}\,s^{-1}$.  
COMPTEL measured the spectrum of Vela at energies of 1 - 30 MeV (Sch\" onfelder et al. 2000) and OSSE at energies between 0.07 -  10 MeV (Strickman et al. 1996).   
Although we have not derived the LAT spectrum for energies below 0.1 GeV, we have compared an extrapolation of the best-fit LAT model spectrum for energies 
$> 0.1$ GeV with the COMPTEL and OSSE data points.   Taking into account only statistical errors, we find that the extrapolated LAT model falls below the COMPTEL 
points but is consistent with the OSSE points.

Abdo et al. (2009a) presented the {\it Fermi} phase-averaged spectrum of Vela using 2 and 1/2 months of data.  Comparing their spectral points with those in Figure 7, the two are consistent within the errors.  The points in Figure 7 are systematically higher for energies below 1 GeV, but this is expected since the more recent IRF gives an increased flux at low energy.    Their spectral parameters for a fit with $b$ fixed to 1.0 are $A = 2.08 \pm 0.04 \pm 0.13 \times 10^{-9}\,\rm cm^2\,s^{-1}\,MeV^{-1}$, $\Gamma = 1.51 \pm 0.01 \pm 0.07$, $E_c = 2.857 \pm 0.089 \pm 0.17$ GeV, and
$H(0.1 - 10 \rm GeV) = (7.87 \pm 0.33 \pm 1.57) \times 10^{-9}\,\rm erg\,cm^{-2}\,s^{-1}$.  In order to compare directly with these parameters, we have fit our 11 month data set assuming $b$ fixed to 1.0 and obtain $A = 2.20 \pm 0.02 \times 10^{-9}\,\rm cm^2\,s^{-1}\,MeV^{-1}$, $\Gamma = 1.57 \pm 0.01$, and $E_c = 3.15 \pm 0.05$ GeV, which are consistent within errors.  The integrated photon flux quoted above is 14\% larger the photon flux derived from the fit reported in Abdo et al. (2009a), which is in agreement with a $\sim$ 25\% increase expected from the differences in the IRFs, partly offset by the different source models and diffuse backgrounds used in the two analyses.  The analysis in Abdo et al. (2009a) was based on pre-launch P6\_V1 IRFs.  The P6\_V3 IRFs (Rando et al. 2009) used here are updated to account for pile-up effects observed in flight data.  Owing to the consequent decreased efficiency for event reconstruction, the effective area is reduced in P6\_V3, especially at lower energies.

\subsubsection{Phase-resolved spectrum}  \label{sec:phase_res}

To explore the phase-resolved spectrum, we used the energy-dependent cut described in \S3.1 to define fixed-count phase bins with 1500 counts each above 0.1 GeV.  A binned maximum likelihood fit was performed in each phase bin, assuming the spectral form in Eqn (\ref{hexp}) with \emph{b} fixed to 1.  For the phase-resolved spectral fits, the same model was used as for the phase-averaged fit, except that the spectral indices of all other sources were held fixed and only the normalizations of the diffuse backgrounds and sources within $5^{\circ}$ of the pulsar were left free in order to ensure that the fits were not overly constrained.  The normalizations of the diffuse backgrounds and sources within $5^{\circ}$ of the pulsar were left free in order to ensure that the fits were not overly constrained.  
It is necessary to choose a fixed value of the $b$ parameter for the phase-resolved spectral fits since allowing $b$ to vary, as was assumed for the phase-averaged analysis, gives an unconstrained fit with errors in $E_c$ and $\Gamma$ of order 100\%. 
We explored different assumptions for $b$, fitting the phase-resolved spectra with
the model in Eqn (\ref{hexp}) with $b$ free to vary and $b$ fixed to 2.  
The log-likelihood ratio test (LRT) prefers a model
with $b$ left free over a model with fixed $b=2$ at about the $3\sigma$
level on average, whereas the model with $b$ left free is not
statistically preferred, on average, over the model with fixed $b=1$.
So, assuming a fixed $b = 1$ for the phase-resolved spectra is statistically justified.
In fact, when comparing the $b$ free to fixed $b = 1$ case the LRT test statistic distribution was similar to that of a $\chi^{2}$ with 1 degree of freedom, 
which is to be expected if the null hypothesis (in this case $b = 1$) is the true hypothesis.

Figures 8 and 9
show the phase-resolved spectral fitting results, plotting the derived photon index $\Gamma$ and 
cutoff energy $E_c$ as a function of phase with the light curve also displayed for reference.  It is apparent that the cutoff energy
varies much more dramatically with phase than the photon index.  The photon index in the peaks is fairly constant, but it increases
outside the peaks (i.e. the spectrum becomes softer) and decreases in the bridge region, becoming hardest between the peaks with a minimum
at the position of P3.  There also seems to be a rough symmetry in the spectral
index variation at a phase about halfway between the two peaks.  This variation in index with phase is similar to and 
consistent with that measured by Kanbach et al. (1994) and by Fierro et al. (1998) using EGRET data, although we see
that with smaller phase bins that the index is roughly constant in the peaks.  For the first time, {\it Fermi} has 
measured the variation of cutoff energy with phase and the result is quite complex.  In contrast to the index 
variation, the pattern is not symmetric.  Generally, the cutoff energy is lowest outside the peaks, increases sharply through the peaks, then falls on the trailing sides of the peaks.  It reaches a maximum of 4 - 4.5 GeV at about the midpoint of
the bridge region and in Peak 2.  These phase regions are in fact correlated with the parts of the light curve that are present at the highest energies.  There is also a minimum in the cutoff energy at about 1.5 GeV in the bridge region at the position of P3 in Figure 9.  This feature is consistent with the shift in P3 with increasing energy to higher phases, where the cutoff energy is higher.  
In Figure 10, we plot the spectral energy distribution in four representative phase bins to illustrate the large variations in spectral shape:
0.133 - 0.135 (in P1), 0.22 - 0.226 (low-energy P3 phase), 0.315 - 0.324 (high-energy 
P3 phase) and 0.562 - 0.563 (in P2).  It is interesting that the spectral index of P3 is nearly 
constant (within the errors) while its cutoff energy increases sharply with phase.

Figures 11 
and 12 show enlargements of the cutoff energy variation in the peaks.  Although the photon index was allowed to float for the 
full phase fits shown in Figures 8 and 9, we find that correlations between index and cutoff energy 
in the fit in neighboring phase bins produce spurious variations.  We therefore have fixed the index through the peaks to their 
weighted mean values, $\Gamma = 1.72 \pm 0.01$ in the range $0.112 \le \phi \le 0.155$ for P1 and $\Gamma = 1.58 \pm 0.01$ 
in the range $0.524 \le \phi \le 0.579$ for P2, to study the cutoff variation alone in the fits shown in Figures 10 and 11. 
The maximum of $E_c$ in P1 occurs at a phase slightly later than the peak in the light curve.

In Figures 8 - 12, we have plotted only the errors due to photon counting statistics.  To explore the possible systematic errors in 
the spectral analysis, we have simulated emission from the Vela pulsar using the \emph{PulsarSpectrum} tool 
(Razzano et al. 2009),
but assuming constant index and cutoff energy with phase.  Performing the same phase-resolved analysis 
as we do for the {\it Fermi} data,
we find a level of variation in the fitted parameters that indicate we should expect deviations on the order of 0.6 GeV on the 
cutoff energy and 0.05 on the photon index just from the fitting behavior.  These systematic variations are of the same order as 
the plotted statistical errors and suggests that any point-to-point fluctuations smaller than about 0.9 GeV in cutoff energy and 0.1 in index, 
such as those seen near phase 0.2 in Fig. 9, are not significant.

As noted above, here is a correlation between the values of $E_c$ and $\Gamma$ in the spectral fits.  To quantify this correlation, we have computed the likelihood on a grid of values for $E_c$ and $\Gamma$ around the best fit values in a few representative phase bins (in P1, P3 region, interpeak and P2) and constructed likelihood error ellipses from these values.  
These error ellipses are not significantly elongated beyond the quoted statistical errors plotted in Figures 8 and 9 and were confined well within the systematic estimates derived from the simulation, quoted above.

The preference for a gradual exponential cutoff ($b < 1$) in the phase-averaged spectral fit is a natural consequence of combining multiple 
simple-exponential-cutoff ($b=1$) spectra with different $E_c$ for each phase bin.
To explore this possibility, we used the \emph{PulsarSpectrum} tool (Razzano et al. 2009) to add simulated phase-varying spectra having $b = 1$ in 100 fixed width phase bins with the same ranges of $E_c$ and $\Gamma$ of the data, and found that the combined spectrum, although not identical to the measured phase-averaged spectrum, indeed is best fit with $b < 1$.
Thus, we have shown that the $b = 0.69$ of the phase-averaged spectrum is
consistent with being a blend of $b = 1$ spectra having different $E_c$.  
Choosing $b = 0.69$ for the phase-resolved analysis would not 
be consistent, since sum of many $b = 0.69$ spectra with different $E_c$ would produce $b < 0.69$ in the phase-averaged
spectrum.

We also used an unfolding method that can derive the spectrum independently of any assumed model (Mazziotta et al. 2009), 
as a check on the  \emph{gtlike} fit
results.  For both the phase-averaged and phase-resolved spectra, we performed the unfolding to derive the spectral points and then
performed  \emph{gtlike} fits to the points, using an on-off method to subtract the background determined by the off-pulse region.  
The parameter values we derive by this method are in agreement with those of the \emph{gtlike} fit.

\subsection{Variability}

Pulsar emission at GeV energies has been found to be stable and non-variable on timescales of weeks to years 
compared to other GeV sources such as blazars.  EGRET studied variability of the Crab, Vela and Geminga over a 3 year 
period (Ramana-Murthy et al. 1995) and found that while the light curves of these pulsars are stable in time, the $> 100$
MeV integral flux of Vela and Geminga showed some variation (the integral flux of the Crab was stable).  A search for 
short-term variability of the Crab pulsed flux at energies $> 50$ MeV (Ramana-Murthy et al. 1998) found no evidence for 
variation on timescales of seconds to hundreds of seconds, nor any correlation with giant pulses.  There was however evidence
for variability in the Crab nebular flux measured by EGRET on a month-year timescale (De Jager et al. 1996).

We looked for flux variability in the Vela pulsar by performing likelihood fits to the data in time intervals between 5 days 
and 1 month and found only modulation at the 55-day precession period of the spacecraft orbit.  Within statistical errors only, 
the fit to the 55-day precession is consistent with the addition of the 
systematic error caused by the variation in effective area due to charged particles during orbital precession.  This 
variation in the LAT effective of area is a known effect that is caused by a change in exposure over the orbital
precession period.  In the case of Vela, the fractional changes in exposure are less than 5 \%.

\subsection{Discussion}

This analysis of the first year of {\it Fermi} LAT data observations of the Vela pulsar provides the highest quality 
measurements to date of pulsar light-curve and spectral characteristics, from which we can begin to study pulsar emission
in more detail.  With increased photon statistics, we 
are able to measure spectral characteristics such as index and cutoff energy in fine phase bins. 
The energy-dependent changes in the light curve are
reflected in the phase-resolved spectrum which shows strong variations in the cutoff energy with phase.  Indeed, 
at phases in the light curve where components disappear or weaken with an increase of energy, such as the outer wings
of the peaks and the full-band position of P3, the cutoff energy is a minimum. 
In examining the off-pulse region (phase $\sim 0.6 - 1.1$) outside the main peaks, we find significant emission above a 
constant background extends beyond the trailing edge of P2 out to phase 0.8.  The pulsed emission apparently only 
turns off between phases 0.8 - 1.0, with pulsed emission present over 80\% of the pulsar period.
The changing P1/P2 ratio is also seen
to result from a lower cutoff energy of P1 (about 2.5 GeV) compared with that of P2 (about 4.5 GeV).  The shape
of the phase-averaged spectrum, having a cutoff that is more gradual than a simple exponential, can 
be understood as a blend of the phase-resolved spectra having a range of cutoff energies and spectral indices convolved with phase-dependent fluxes.  
In fact, the $E_c \simeq 1.3$ 
GeV value derived for the phase-averaged fit with floating $b$ is close to the lowest $E_c \sim 1.4$ in the peaks of the phase-resolved
spectra (where the highest flux is measured).
The detailed behavior of $E_c$ with phase is quite complex and to fully understand its meaning may require 
detailed modeling as well as 
comparison with phase-resolved spectra of other pulsars with a variety of light curve shapes.

We nevertheless have learned, and can continue to learn, a great deal about emission models with Vela data alone.  
The question of whether the pulsed emission originates near the neutron star surface or in the outer
magnetosphere has already been settled in favor of the outer magnetosphere.   The measurement of the phase-averaged 
spectral cutoff shape (Abdo et al. 2009a) excluded a super-exponential turnover, a hallmark of magnetic
pair creation attenuation at low altitudes, at high
levels of significance.  The wide pulse profile that covers a large fraction of solid angle of the 
sky as well as the phase lag of the $\gamma$-ray peaks with the radio peak also favors
interpretation in outer magnetosphere models such as the outer gap (OG) (Cheng, Ho \& Ruderman 1986, 
Romani \& Yadigaroglu 1995), slot gap (SG) (Muslimov \& Harding 2004) or two-pole caustic model (TPC) (Dyks \& Rudak 2003).  
These models both have a large flux correction factor, $f_{\Omega} \sim 1$ (Watters et al 2009), which is needed to convert the 
phase-averaged energy flux we observed in the light curve to the total radiated luminosity, 
\be
L_{\gamma} = 4\pi d^2 f_\Omega H,
\ee
where $d$ is the source distance.  The correction factor $f_{\Omega}$ is model dependent and is a function of the
magnetic inclination angle $\alpha$ and observer angle $\zeta$ with respect to the rotation axis.  For Vela, we can
obtain a good estimate of $\zeta$ from modeling the pulsar wind nebula, as observed by Chandra. According to 
Ng \& Romani (2008), the derived tilt angle of the torus gives $\zeta = 64^{\circ}$.  Using the estimate of
gap width from Watters et al. (2009), $w \simeq (10^{33}\,{\rm erg\,s^{-1}}/\dot E_{sd})^{1/2}$, $w = 0.01$ for 
Vela, where $\dot E_{sd} = 6.3 \times 10^{36}\,\rm erg\,s^{-1}$.   This $\zeta$ value, and
the observed peak separation of 0.42 constrains the inclination angle to $\alpha = 75^\circ$ 
for the OG model,
giving $f_{\Omega} = 1.0$, and $\alpha = 62^\circ - 68^\circ$ for the SG/TPC model, giving $f_{\Omega} = 1.1$.
The best values derived from fits of the radio polarization position angle vs. phase are $\alpha = 53^\circ$ and $\beta = 6.5^\circ$ (Johnston et al. 2005). This gives $\zeta = \alpha + \beta = 59.5^\circ$, not too far from the torus value.
However, the $\alpha$ is significantly lower than either the OG or the SG/TPC would predict, but closer to the SG/TPC range.
Assuming that $f_\Omega \sim 1$ and the 
distance of $d = 287 ^{+19}_{-17}$ pc for Vela, we can estimate the total luminosity, $L_\gamma = 8.2 ^{+1.1}_{-0.9} 
\times 10^{34}\,
f_{\Omega}\rm erg\,s^{-1}$ and pulsed efficiency as $\eta_\gamma = L_\gamma/\dot E = 0.01 \,f_{\Omega}$.

We have seen from the phase-resolved spectral analysis that both the photon index and cutoff energy vary with 
pulse phase.  In all phase intervals, the spectra are well fit with a power law and simple exponential cutoff
form.  This spectral form is characteristic of a number of non-thermal radiation mechanisms, such as curvature, 
synchrotron and inverse Compton emission from relativistic electrons either at a single energy or having a power-law
spectrum with a high-energy cutoff.  In all cases, as noted above, it indicates that the emission is originating
at least several stellar radii above the neutron star surface, where the magnetic field strength has dropped to 
values too low (assuming a dipolar field geometry) for magnetic pair attenuation to operate. 
The minimum emission radius can be estimated to be where
    photons at that location have an optical depth of unity to
    single-photon pair creation $\gamma\to e^+e^-$ and can be derived from Eq. [1] of Baring (2004) (see also the
discussion in Abdo et al. 2009a).  Adopting a value of around $3E_c$ for the maximum emission energy,   
$\epsilon_{\rm max}=7.5$ GeV in P1 ($E_c\sim 2.5$GeV; see Figure 9) yields
    $r\gtrsim 3.0 R_{\ast}$. In P2, the higher energy choice of
    $\epsilon_{\rm max}=12.0$ GeV ($E_c\sim 4.0$GeV) derives the bound
    $r\gtrsim 3.5 R_{\ast}$.  These altitudes are slightly higher
    than that obtained in the first {\it Fermi} paper on the Vela pulsar
    (Abdo et al. 2009a) because the pulse-phase statistics
    have improved deep into the high-energy turnover as photon counts have
    accumulated.  

For the case of curvature radiation (CR), the spectrum of a mono-energetic electron distribution has a photon index 
of 2/3 and a cutoff energy (in $mc^2$ units) of 
$E_c = 1.5 (\lambar/\rho_c) \gamma^3 = 0.25\, {\rm  GeV} \gamma_7^3/\rho_8 $, where $\lambda_c = 2\pi \lambar$ is the
electron Compton wavelength, 
$\rho_8 \equiv \rho_c/10^8$ cm is the local field line radius of curvature and $\gamma_7 \equiv \gamma/10^7$ 
is the electron Lorentz factor.  For synchrotron radiation (SR) from
mono-energetic electrons, the spectrum is a power law with index 2/3 with a high-energy cutoff of $E_c = 1.5 B'\gamma^2
\sin\theta $, where $B' \equiv B/4.4 \times 10^{13}$ Gauss is the local magnetic field strength and 
$\theta$ is the particle pitch angle.  
In both of these cases, the spectrum is steeper if the
radiating electrons have a power law distribution of Lorentz factors.  The inverse Compton (IC) spectrum has a cutoff
at $E_c =$ min$[\gamma^2 \epsilon, \gamma]$, where $\epsilon$ is the maximum energy of the soft photons.  Generally, CR
and SR will dominate over IC in pulsar magnetospheres, even though the IC spectrum can extend to much higher energies
(even to the TeV range).  In nearly all current pulsar outer-magnetosphere emission models, 
the spectrum above 100 MeV is CR from nearly 
mono-energetic or power-law electrons, since it is difficult to maintain high enough pitch angles of the electrons
to radiate to GeV energies in the relatively low magnetic fields in the outer magnetosphere.  The Lorentz factors of
the CR emitting electrons, which are continuously being accelerated by the electric field $E_\parallel$ parallel to the
magnetic field, are limited
by CR reaction so that there is force balance that maintains a steady-state Lorentz factor,
\be  \label{eq:gammaCR}
\gamma_{\sc{CR}} = \left[{3\over 2}{E_{\parallel} \rho_c^2\over e} \right]^{1/4}. 
\ee
In all the outer magnetosphere models such as the SG (Muslimov \& Harding 2004) or OG (Zhang et al. 2004, Hirotani 2008), 
$E_\parallel \simeq C(r) B_{\rm LC} w^2$, where $B_{\rm LC}$ is the magnetic field strength 
at the light cylinder, $w$ is the gap width and $C(r)$ is some function of the emission radius.  For these models,
the values of $E_\parallel$ turn out to be fairly similar for the same gap width and the steady-state electron
Lorentz factors are all around $\gamma_{\sc{CR}} \sim (2-3) \times 10^7$.  The resulting cutoff energy (in $mc^2$), 
\be  \label{eq:Ecr}
E_{CR} = {3\over 2} {\lambar\over \rho_c} \gamma_{\sc{CR}}^3 = 0.32\,\lambda_c \left({E_\parallel \over e} \right)^{3/4} \rho_c^{1/2} 
\ee
thus falls in the 1-5 GeV energy range, consistent with that measured for pulsar spectra (Abdo et al. 2009c) 
assuming that $\rho_c \sim (0.1-0.6) R_{\rm LC}$ in the outer magnetosphere (e.g. Zhang et al. 2004).  

The dependence of $E_\parallel$ on $\rho_c$ and radius $r$ can be constrained from the spectral 
measurements if the emission from Vela is primarily CR at {\it Fermi} energies.  The single electron
CR spectrum can be written approximately as (e.g. Harding et al. 2008)
\be
{dN_{CR}(E) \over dE} \simeq {\alpha c\over (\lambar^{1/3}\rho_c^{2/3})}  \,E^{-2/3}\exp(-E/E_{\rm CR}),
\ee
so that the level of the spectrum below the cutoff depends only on $\rho_c$ and is a maximum for minimum $\rho_c$.
Assuming that the phase-averaged spectrum is the sum of phase-resolved spectra, and using the expression for the cutoff
energy in Eqn (\ref{eq:Ecr}) assuming $E_c = E_{\rm CR}$, 
we would expect that the effective cutoff energy of the phase-averaged spectrum would
be close to the minimum value of $E_c$ of the phase-resolved spectra if $E_c \propto \rho_c^n$ where $n$ is positive.
On the other hand, if $n$ is negative, then we would expect that the effective phase-averaged cutoff energy would
be closer to the maximum value of $E_c$.  In the peaks, $E_c$ varies between about 1.4 and 5 GeV.  
Since the $E_c = 1.3$ GeV measured for the phase-averaged spectrum is closer to
the minimum value $E_c \sim 1.4$ GeV in the peaks, we deduce that $n$ is positive.  This constrains
$E_\parallel \propto \rho_c^m$, where $m > -2/3$.  If $\rho_c \sim (r\,R_{LC})^{1/2}$, where $R_{LC}$ is the light cylinder 
radius (Zhang et al. 2004), then  $E_\parallel \propto r^q$,
where $q > -1/3$, so that any decrease of $E_\parallel$ with radius is constrained to be very gradual.

In outer magnetosphere models, the emission at different phases of the light curve originate from different ranges of 
emission radii.  The pulse phase therefore maps emission altitude in a complicated way.  
The peaks are caustics that result from phase shifts due to relativistic aberration and retardation 
that nearly cancel those due to the dipole field line curvature on trailing field lines (Morini 1983).  The emission
from a large range of altitudes will arrive at a small range of phase in the light curve, producing a sharp peak. 
In two-pole caustic models (Dyks \& Rudak 2003) (including the SG), emission extends from the neutron star surface to near
the light cylinder, the caustics from both magnetic poles can be
observed and the two peaks in Vela-like light curves are the caustics from two poles.  In OG models (Romani \&
Yadigaroglu 1995), the gap exists only above the null charge surface ($\zeta = 90^\circ$) so that the trailing caustic 
from only one pole can be observed and forms the second peak.  The first peak is a caustic caused by 
field lines from the same pole that overlap near the light cylinder.  
This first OG peak is also present in TPC/SG models, where field lines from both poles overlap at this phase.

To what degree do the detailed characteristics of the light curve and phase-resolved spectrum challenge the current emission models?  The key to understanding variations of flux and spectrum with phase is likely to be found in the large variations with pulse phase of emission radii and field line radius of curvature in TPC (Dyks et al. 2004) and OG (Romani 1996, Cheng et al. 2000, Takata \& Chang 2007) geometries.  Thus, the variations in $E_c$ with pulse phase that we see for Vela could be a result of CR from different emission altitudes having different $E_c$ and $\rho_c$.  In both models, the minimum emission radius and radius of curvature increase on the leading edge of the P1 and decreases on the trailing edge of P2.  The observed decrease in width of P1 and P2 with increasing energy,  mostly on the outer edges could thus be understood in both TPC and OG models if $E_c$ depends on a positive power of curvature radius.  However, any emission at the outer edges of the peaks requires extension of the OG below the null surface, so that the original OG geometry has been ruled out in favor of the revised OG models of Hirotani (2006) and Takata \& Chang 2007).  Furthermore, the observed extension of the trailing edge of P2 to phase 0.8 (Figure 6) indicates the presence of emission far below the null surface, albeit at relatively low levels.  TPC/SG models predict some level of pulsed emission at off-pulse phases (Dyks et al. 2004), and thus could account more naturally for the extended off-pulse emission we observed in the Vela light curve.  The observed decrease in the P1/P2 ratio with energy is a consequence of the maximum $E_c$ being lower for P1 than for P2.  Indeed, this is true for both TPC/SG and OG models where the mean emission altitude of P1 is higher than that of P2.  The possible 
splitting of P2 into two smaller peaks above $\sim 8$ GeV, although marginally detected at this point, will provide an
additional model constraint.  Interestingly, double P2 peaks appear in geometric OG model light curves (Watters et al. 2009),
but full radiation models are required to address the observed energy dependence of this new feature.

The presence of a P3 component is natural in TPC models, where it has the same origin as the first OG peak (i.e. emission of overlapping field lines near the light cylinder) and appears as a trailing shoulder on P1 (Dyks et al. 2004).  
The phase shift of P3 could be explained in the TPC geometry if
the $E_{\parallel}$ decreases with altitude along the field lines.  In this model, the
radius of emission at the low-energy phase of P3 is at the light cylinder and
decreases with increasing phase.  Therefore, the $E_{\parallel}$ (and thus $E_c$ according
to Eqn (5)) is increasing between phase 0.2 and 0.35 (as in Fig. 9).  That would
cause P3 to move from phase 0.2 (with lower $E_c$) to phase 0.35 (with higher $E_c$)
with increasing energy.  A P3 component is not naturally produced in geometric OG models, so such a component would have to result from an additional particle population or emission component.  Ultimately, understanding how the model variations in emission radius and curvature radius map to the observed $E_c$ variations and other light curve variations will require more detailed modeling, including the acceleration and radiation physics.  Thus, we have attempted only a qualitative discussion here, but emphasize the potential power of such future modeling.  A particular challenge for  
models is to reproduce both the energy-dependent light curves and the phase-resolved spectra.
One open question is, how do variations in $E_\parallel$ and curvature radius with altitude combine to determine the $E_c$ at each phase?  Since the complex radius of curvature variations depend on the global magnetic field structure, the phase-resolved spectral results presented here for the Vela pulsar, as well as those of other bright pulsars, have the potential to constrain the global field geometry.  
\vskip 0.2cm

The \textit{Fermi} LAT Collaboration acknowledges generous ongoing support
from a number of agencies and institutes that have supported both the
development and the operation of the LAT as well as scientific data analysis.
These include the National Aeronautics and Space Administration and the
Department of Energy in the United States, the Commissariat \`a l'Energie Atomique
and the Centre National de la Recherche Scientifique / Institut National de Physique
Nucl\'eaire et de Physique des Particules in France, the Agenzia Spaziale Italiana
and the Istituto Nazionale di Fisica Nucleare in Italy, the Ministry of Education,
Culture, Sports, Science and Technology (MEXT), High Energy Accelerator Research
Organization (KEK) and Japan Aerospace Exploration Agency (JAXA) in Japan, and
the K.~A.~Wallenberg Foundation, the Swedish Research Council and the
Swedish National Space Board in Sweden.

Additional support for science analysis during the operations phase is gratefully
acknowledged from the Istituto Nazionale di Astrofisica in Italy and the Centre National d'\'Etudes Spatiales in France.

\noindent

\acknowledgments  

\newpage
\begin{table}
\caption{Light-Curve Fit Paramters} \label{tbl-1}
\begin{center}
{\tiny
\begin{tabular}{lccccccc}
\tableline
Energy (GeV) & 0.02-0.1 & 0.1-0.3 & 0.3-1.0 & 1.0-3.0 & 3.0-8.0 & 8.0-20.0 & 0.02 - 20.0 \\
\tableline
P1 Phase & 
0.1308 $\pm$ 0.0013 &  0.1307 $\pm$ 0.0003 &  0.1310 $\pm$ 0.0002&  0.1306 $\pm$ 0.0003&  0.1303 $\pm$ 0.0008 &  0.1349 $\pm$ 0.0045 & $0.13128 \pm 0.00016$\\
P2 Phase &
0.5617 $\pm$ 0.0035& 0.5642 $\pm$ 0.0007& 0.5651 $\pm$ 0.0005& 0.5675 $\pm$ 0.0007& 0.5696 $\pm$ 0.0013& 0.5699 $\pm$ 0.0007 & $0.56513 \pm 0.00022$\\
P3 Phase &
0.175 $\pm$ 0.015& 0.1932 $\pm$ 0.0026& 0.2103 $\pm$ 0.0020& 0.2544 $\pm$ 0.0024& 0.2901 $\pm$ 0.0027& 0.3034 $\pm$ 0.0023 & $0.2167 \pm 0.0015$\\
P1 Inner Width & 
0.0093 $\pm$ 0.0022 & 0.0112 $\pm$ 0.0005 & 0.0111 $\pm$ 0.0004 & 0.0161 $\pm$ 0.0009 & 0.0263 $\pm$ 0.0029 & 0.0099 $\pm$ 0.0047  & $0.01165 \pm 0.00030$\\
P1 Outer Width & 
0.0098  $\pm$ 0.0021& 0.0098 $\pm$ 0.0004& 0.0088 $\pm$ 0.0003& 0.0074 $\pm$ 0.0002& 0.0049 $\pm$ 0.0007& 0.0022 $\pm$  0.0008 & $0.00939 \pm  0.00021$\\
P2 Inner Width & 
0.0263 $\pm$ 0.0035& 0.0326 $\pm$ 0.0007& 0.0318 $\pm$ 0.0004 & 0.0327 $\pm$  0.0007 & 0.0329 $\pm$ 0.0013& 0.0256 $\pm$ 0.0006 & $0.03140 \pm  0.00033$\\
P2 Outer Width &  
0.0170 $\pm$ 0.0027& 0.0159 $\pm$ 0.0004& 0.0121 $\pm$ 0.0002& 0.0076 $\pm$ 0.0002& 0.0044 $\pm$ 0.0004& 0.0027 $\pm$  0.0003 & $0.01188 \pm 0.00019$\\
P3 Width & 
0.340 $\pm$ 0.043& 0.4052 $\pm$ 0.0086 &  0.3994 $\pm$ 0.0066 &  0.3082 $\pm$ 0.0086 &  0.2062 $\pm$ 0.0096 &  0.1298 $\pm$ 0.0049 & $0.3984 \pm 0.0052$\\

\tableline
\end{tabular}
}
\end{center}
\end{table}

\newpage
\begin{figure}  
\includegraphics[width=160mm]{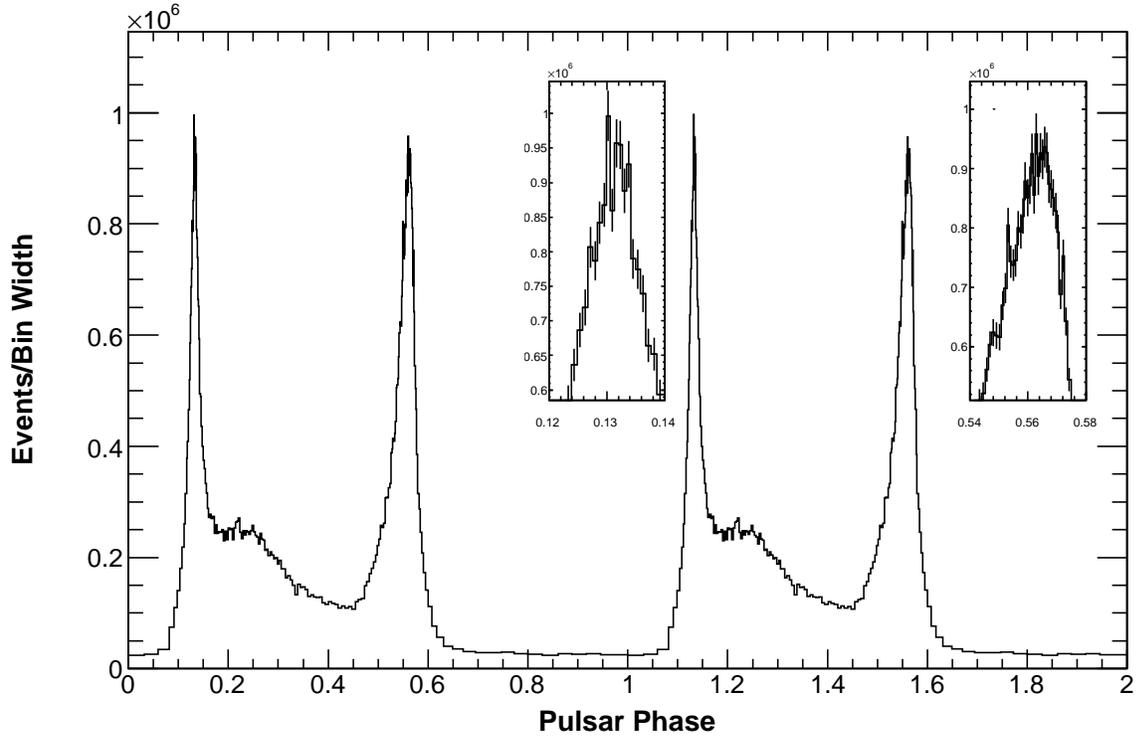}
\vskip 1.0cm
\caption{Light curve for energy range 0.02 - 300 GeV for 203 fixed-count phase bins with 750 photons per bin. Insets show details of P1 and P2.}
\label{fig:LCfull}
\end{figure}

\newpage
\begin{figure}  \label{fig:LCenergy}
\includegraphics[width=160mm]{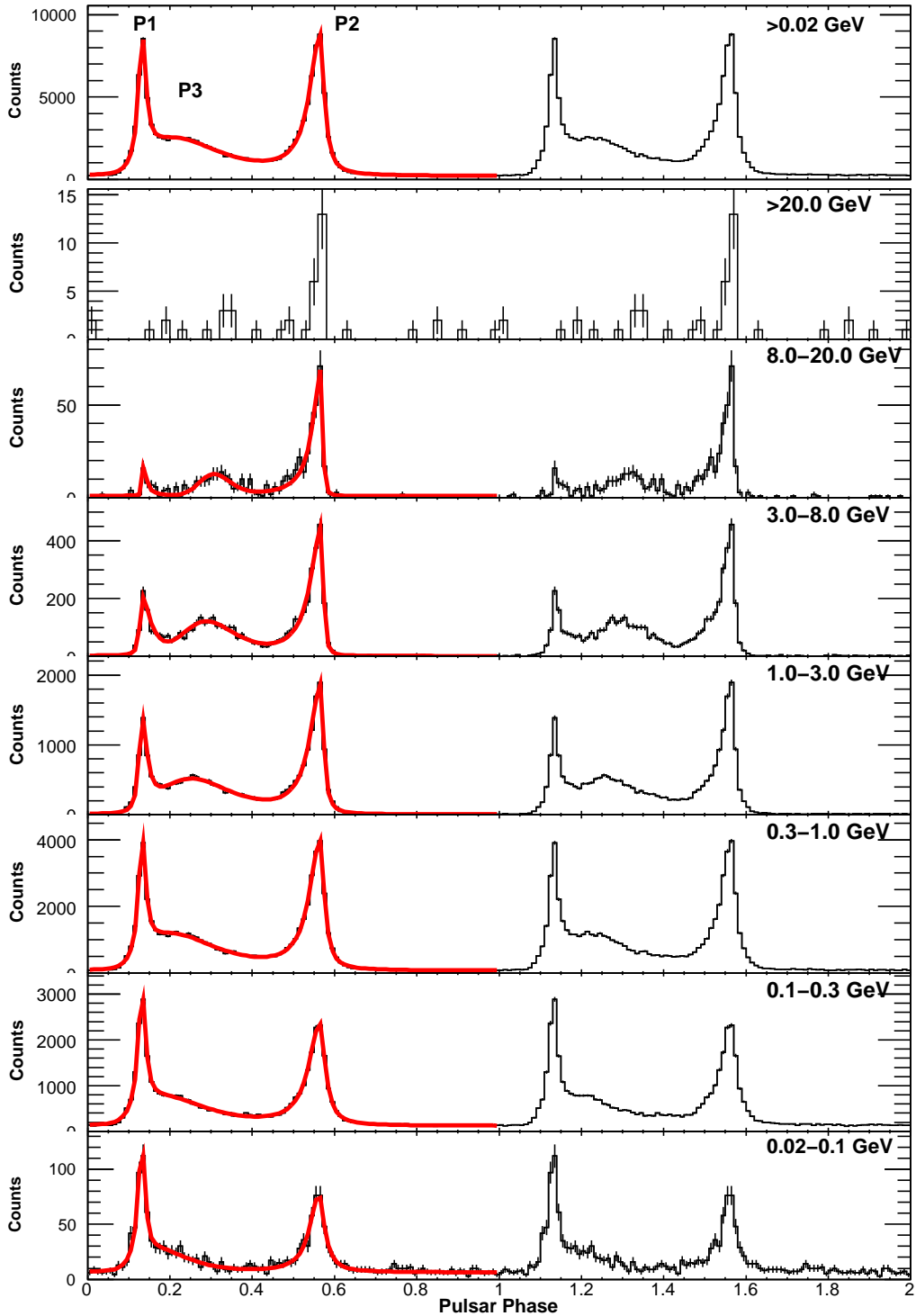}
\vskip 1.0cm
\caption{Light curves for different energy bands over two pulse periods.  Joints fits of P1,P2 and P3 in each energy band (as described in the text) are superposed on the light curves in the first cycle.}
\end{figure}

\begin{figure}  \label{fig:P1P2Widths}
\includegraphics[width=160mm]{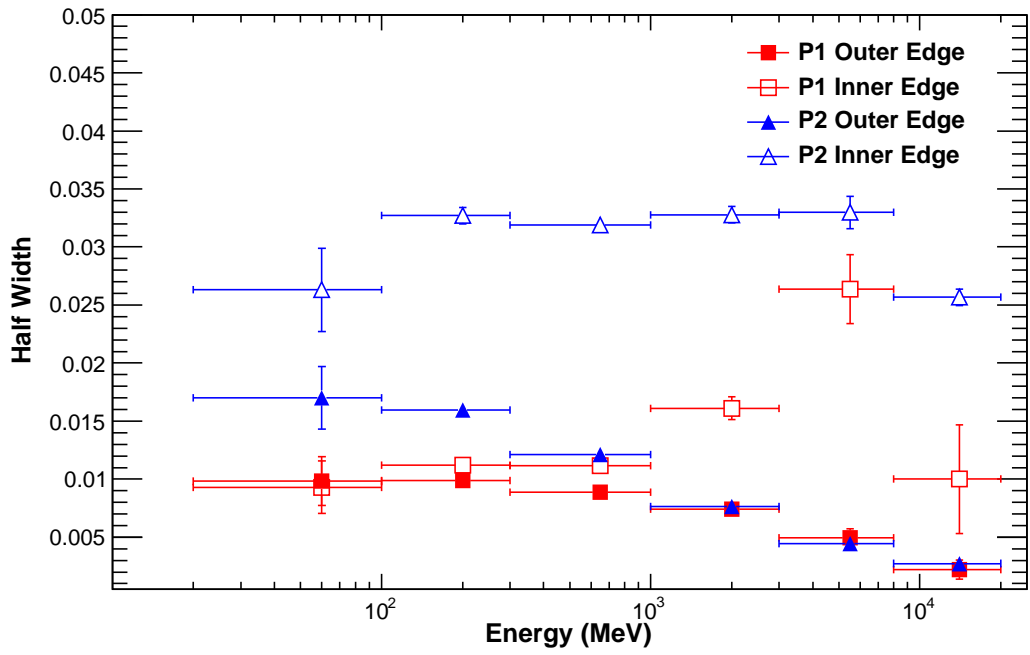}
\vskip 1.0cm
\caption{Measured HWHM of P1 and P2 from two-sided Lorentzian fits as a function of energy.}
\end{figure}

\newpage
\begin{figure}  \label{fig:P3Width}
\includegraphics[width=160mm]{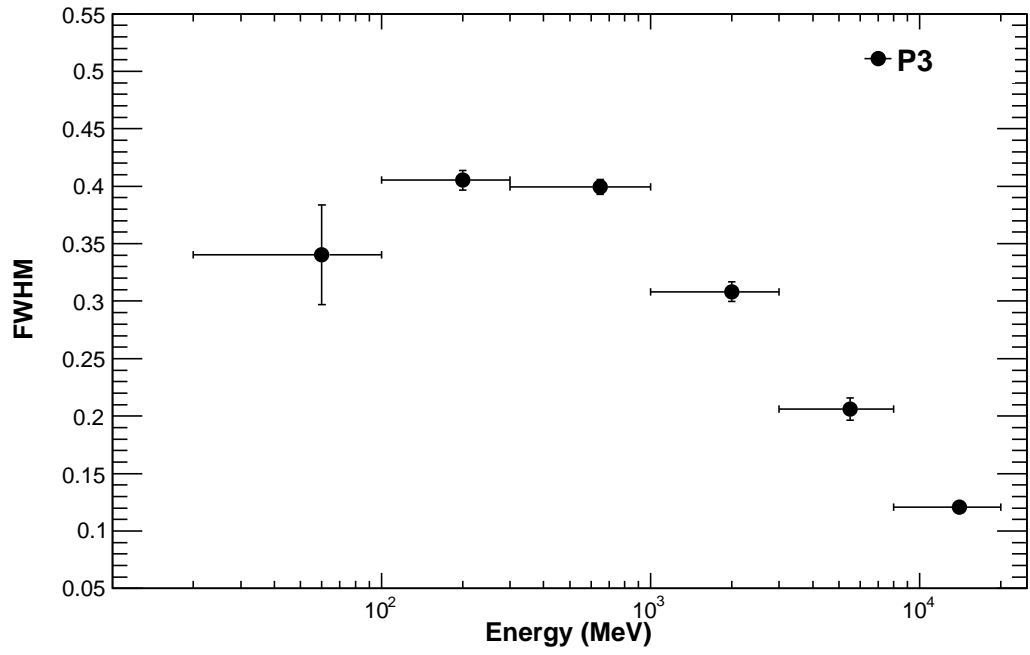}
\vskip 1.0cm
\caption{Measured width (FWHM) of P3 as a function of energy.}
\end{figure}

\newpage
\begin{figure}  \label{fig:LCcontour}
\vskip -4.1cm
\includegraphics[width=160mm]{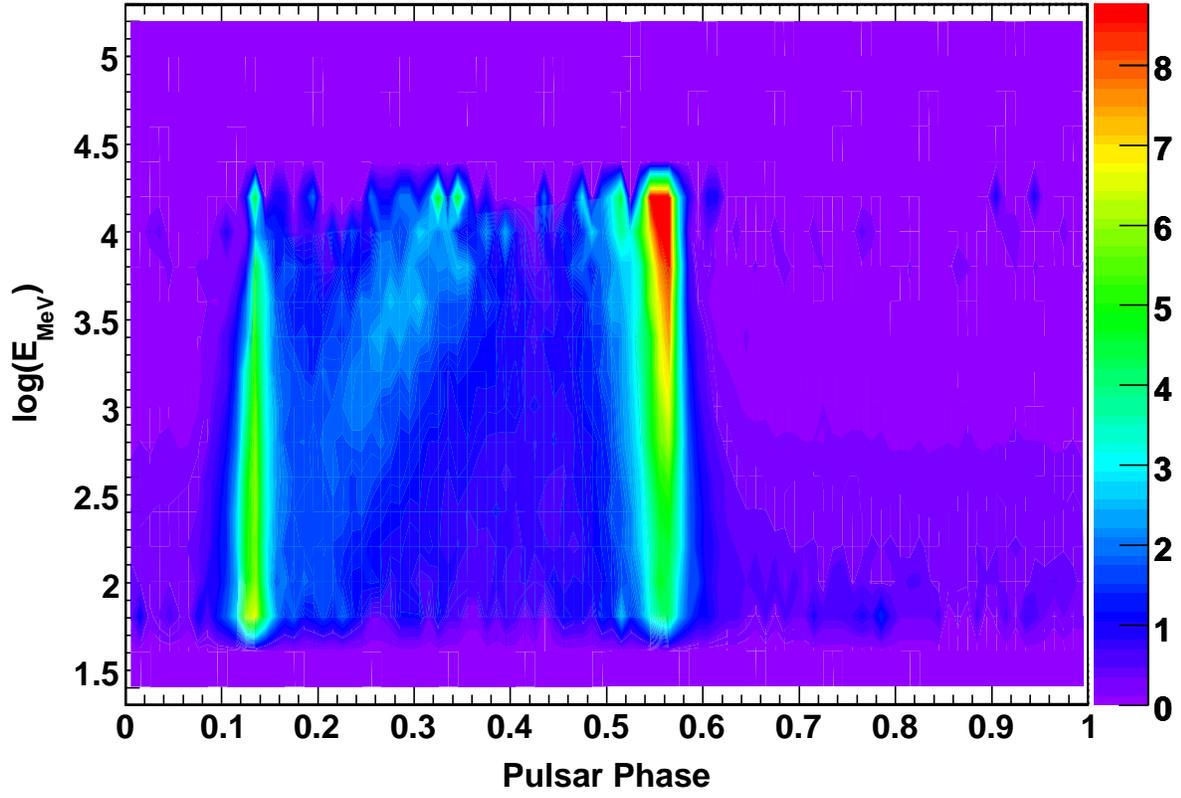}
\vskip 1.0cm
\caption{Plot of pulse profile as a function of energy.  The color scale is relative counts in each bin on a linear scale.}
\end{figure}

\newpage
\begin{figure}  \label{fig:LC_offPulse}
\includegraphics[width=160mm]{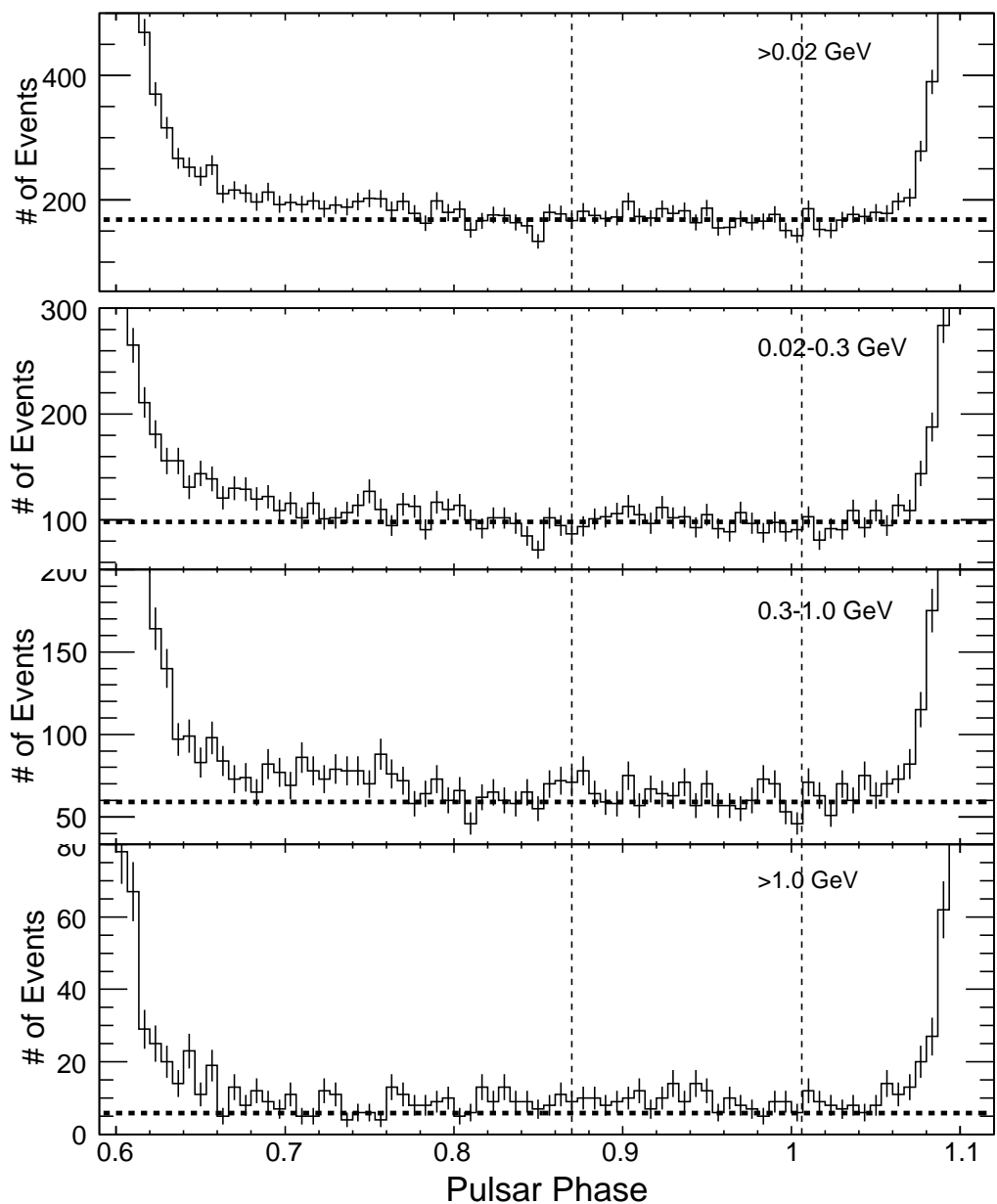}
\vskip 1.0cm
\caption{Detail of the off-pulse phase region for 150 equal-size phase bins for the full energy range 0.02 - 300 GeV and in three different energy bands.  The horizontal dashed line denotes the background level determined from a simulation (see \S 3.2.2). The dashed vertical lines mark the phases of the RXTE P3 and P4 (Harding et al. 2002).}
\end{figure}

\newpage
\begin{figure}  \label{fig:ave_spec}
\includegraphics[width=160mm]{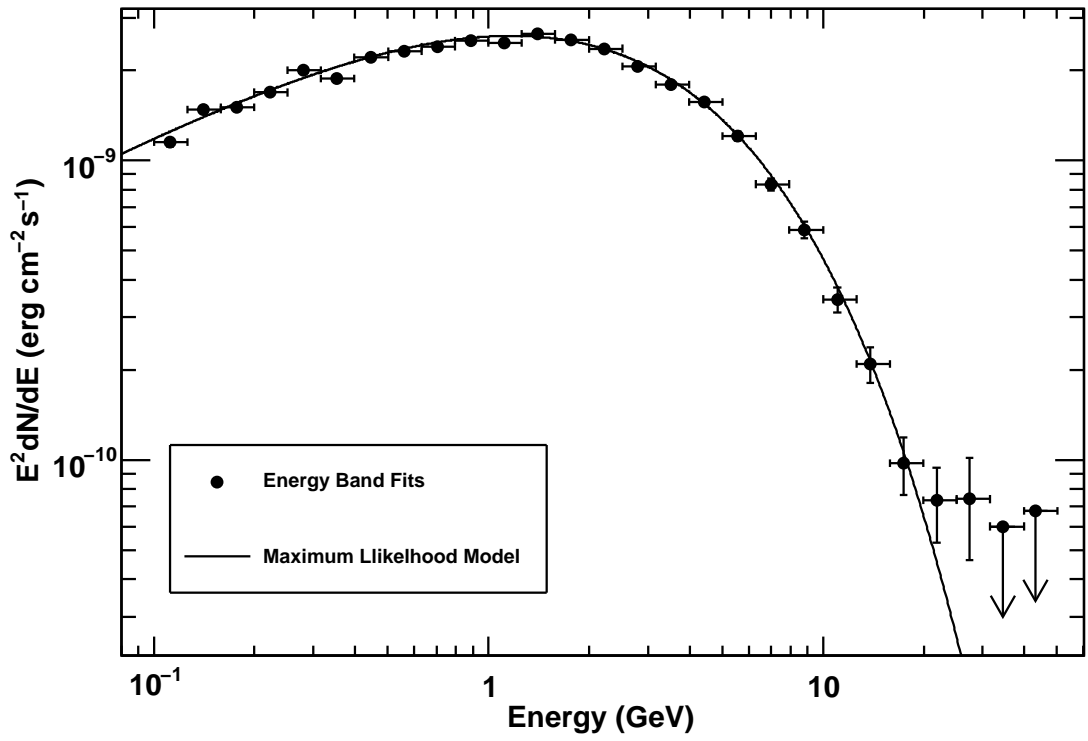}
\vskip 1.0cm
\caption{Phase-averaged spectrum for $E > 0.1 - 100 $ GeV.}
\end{figure}

\newpage
\begin{figure}  \label{fig:pres_spec_index}
\includegraphics[width=160mm]{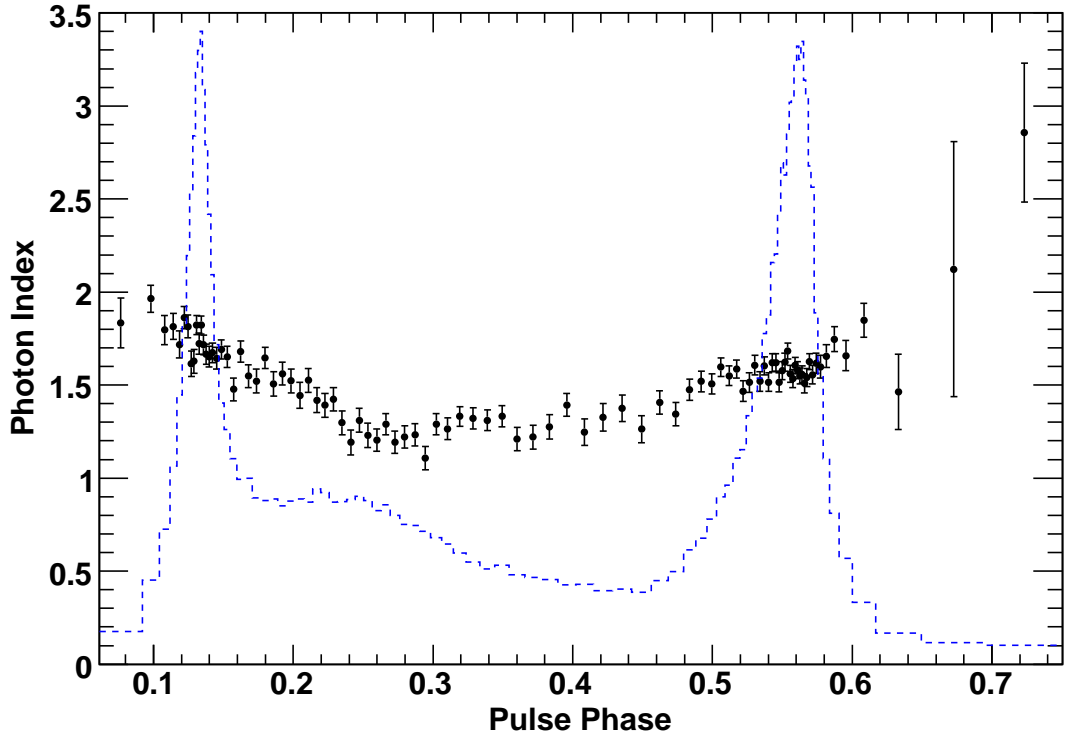}
\vskip 1.0cm
\caption{Photon index vs. phase from fits in fixed-count phase bins of 1500 photon counts per bin. The 
error bars denote statistical errors.}
\end{figure}

\newpage
\begin{figure}  \label{fig:pres_spec_Ec}
\includegraphics[width=160mm]{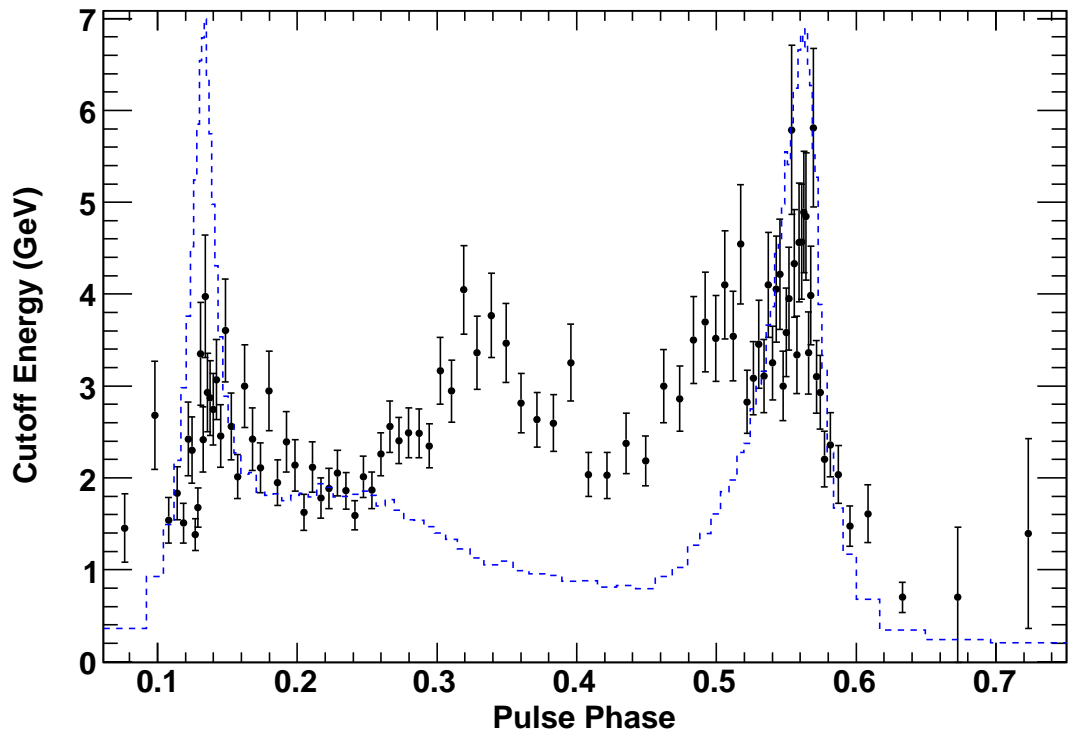}
\vskip 1.0cm
\caption{Spectral cutoff energy vs. phase from fits in fixed-count phase bins of 1500 photon counts per bin.  The solid 
error bars denote statistical errors.}
\end{figure}

\newpage
\begin{figure}  \label{fig:Ec_gamconst_p1}
\vskip -7.0cm
\includegraphics[width=180mm]{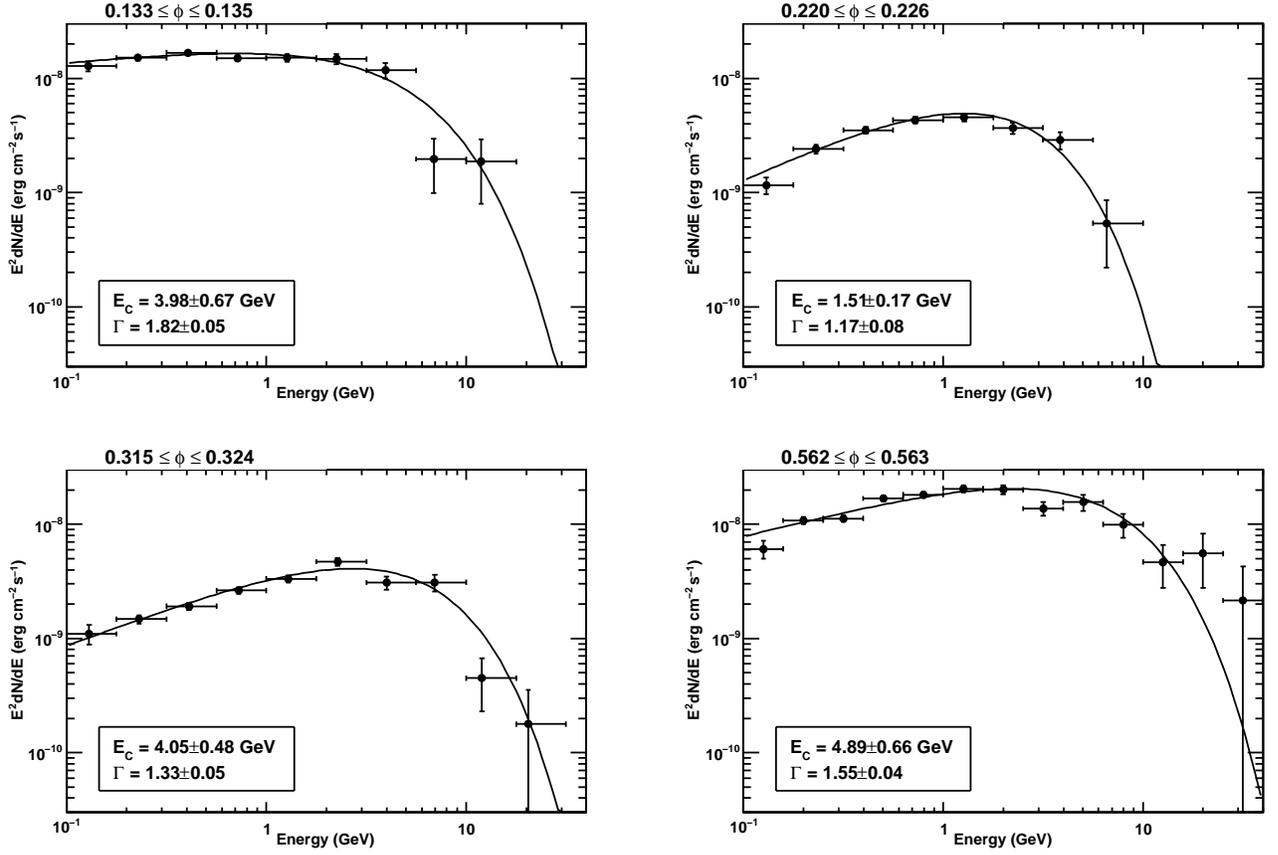}
\vskip 1.0cm
\caption{Spectral energy distribution in four phase intervals of Figure 9.  The individual spectra have been exposure corrected to account for the fact that the fitting was done in a small phase bin.}
\end{figure}

\newpage
\begin{figure}  \label{fig:Ec_gamconst_p1}
\includegraphics[width=160mm]{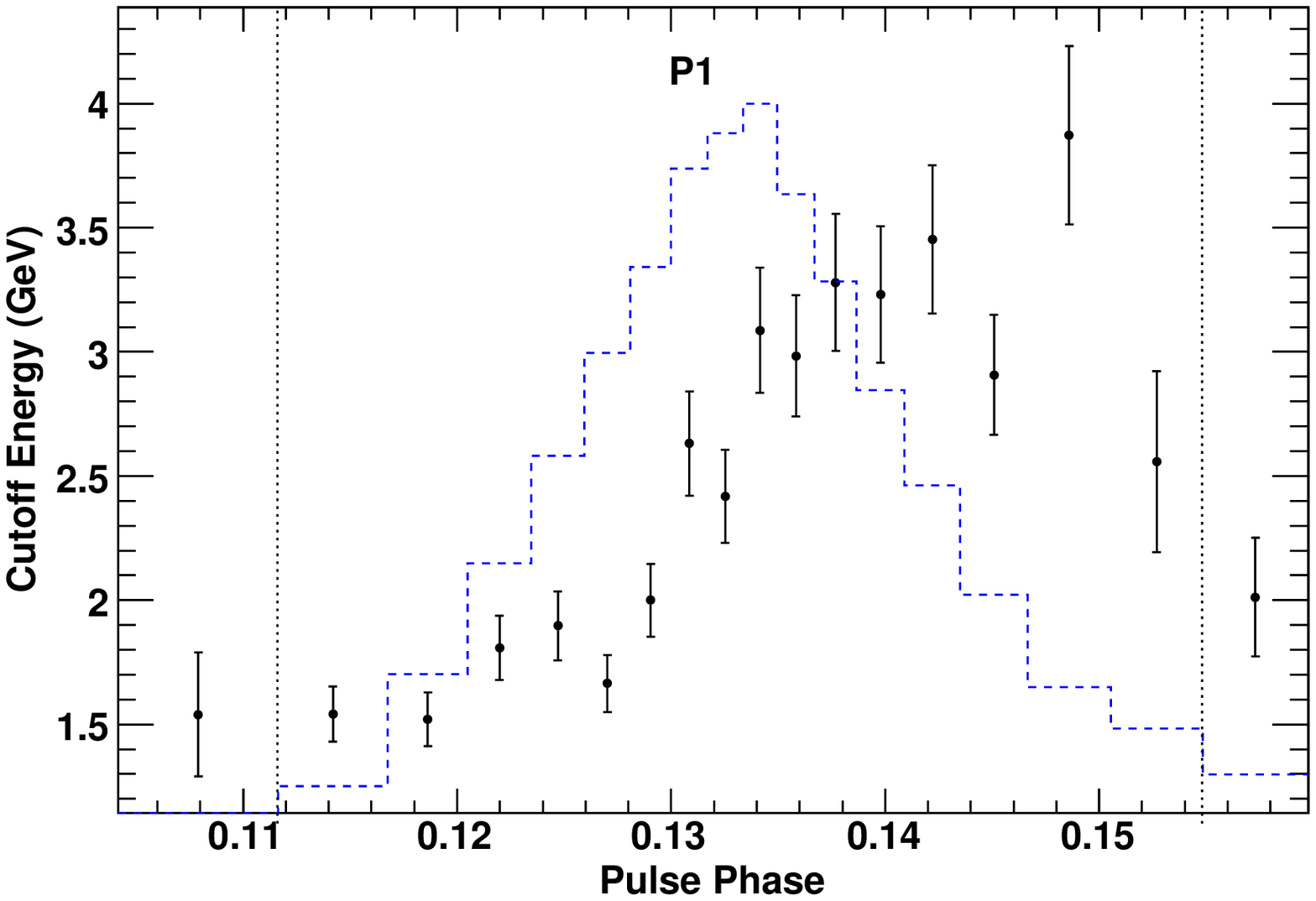}
\vskip 1.0cm
\caption{Spectral cutoff energy vs. phase in P1 from fits in fixed-count phase bins of 1500 photon counts per bin, where the
photon index was held fixed to a value of $1.72 \pm 0.01$ over the phase range between the dotted lines.}
\end{figure}

\newpage
\begin{figure}  \label{fig:Ec_gamconst_p2}
\includegraphics[width=160mm]{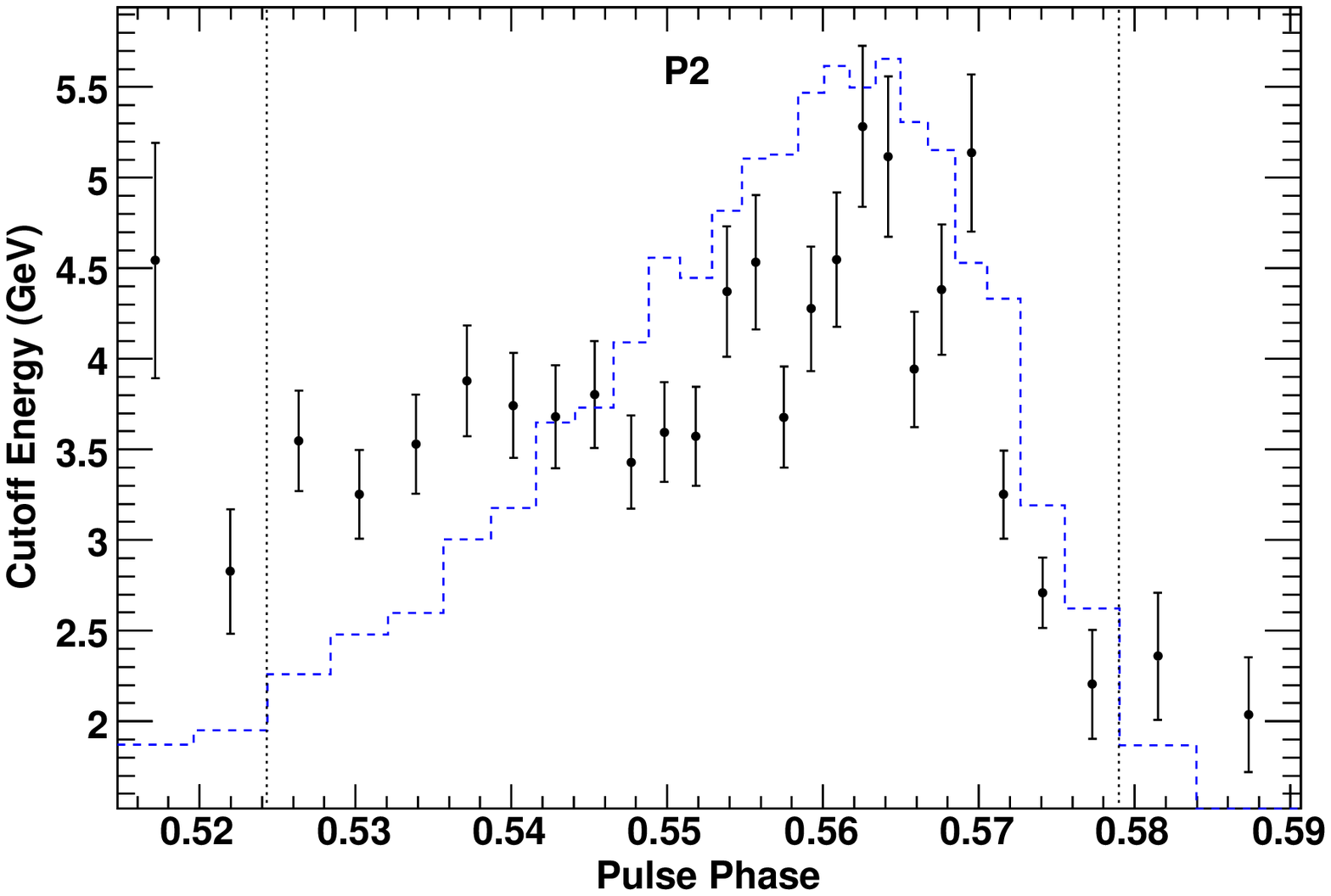}
\vskip 1.0cm
\caption{Spectral cutoff energy vs. phase in P2 from fits in fixed-count phase bins of 1500 photon counts per bin, where the
photon index was held fixed to a value of $1.58 \pm 0.01$ over the phase range between the dotted lines.}
\end{figure}

\end{document}